\documentclass[aps,prb,twocolumn,superscriptaddress,amsmath,amssymb]{revtex4-1}

\usepackage{xcolor}
\usepackage{graphicx}
\usepackage{bm}
\usepackage{lineno}
\usepackage{multirow}
\usepackage{footnote}
\usepackage{soul}
\usepackage[colorinlistoftodos]{todonotes}

\newcommand{\Pblue}{P$_{\rm blue}$}
\begin{document}

\title{Strain and electronic properties at the van der Waals interface of phosphorus/boron nitride heterobilayers}

\author{Baptiste Bienvenu}
\affiliation{Laboratoire d'Etude des Microstructures, ONERA-CNRS, UMR104, Universit\'e Paris-Saclay, BP 72, 92322 Ch\^atillon Cedex, France}
\affiliation{DEN - Service de Recherches de M\'{e}tallurgie Physique (SRMP), CEA Saclay, France}
\author{Hakim Amara}
\address{Laboratoire d'Etude des Microstructures, ONERA-CNRS, UMR104, Universit\'e Paris-Saclay, BP 72, 92322 Ch\^atillon Cedex, France}
\address{Universit\'{e} de Paris, Laboratoire Materiaux et Phenomenes Quantiques, CNRS, F-75013, Paris, France}
\author{Fran\c{c}ois Ducastelle} 
\address{Laboratoire d'Etude des Microstructures, ONERA-CNRS, UMR104, Universit\'e Paris-Saclay, BP 72, 92322 Ch\^atillon Cedex, France}
\author{Lorenzo Sponza}
\address{Laboratoire d'Etude des Microstructures, ONERA-CNRS, UMR104, Universit\'e Paris-Saclay, BP 72, 92322 Ch\^atillon Cedex, France}

\begin{abstract}
We study the mechanical and electronic properties of heterobilayers composed of black phosphorus (BP) on hexagonal boron nitride (hBN) and of blue phosphorus (\Pblue) on hBN by means of ab-intio density functional theory.
Emphasis is put on how the stress applied on the constituent layers impact their structural and electronic properties.
For this purpose, we adopt a specific scheme of structural relaxation which allows us to distinguish between the energy cost of distorting each layer and the gain in stacking them together. In most cases we find that the BP tends to contract along the softer armchair direction, as already reported for similar structures. This contraction can attain up to 5\% of strain, which might deteriorate its very good transport properties along the armchair direction. To prevent this, we propose a twisted-bilayer configuration where the largest part of the stress applies on the zigzag axis, resulting in a lower impact on the transport properties of BP.
We also investigated a \Pblue/hBN bilayer. A peculiar hybridization between the valence states of the two layers lets us suggest that electron-hole pairs excited in the bilayer will exhibit a mixed character, with electrons localized solely in the \Pblue{ }layer, and holes spread onto the two layers.

\end{abstract}

\date{\today}

\maketitle


Past years saw a massive investigation of atomically thin materials, or 2D materials, demonstrating both theoretically and experimentally many of their interesting electronic and mechanical properties~\cite{novoselov_nature2012, Xia-natpho2014, li_natnanotech2014, liu_acsnano2014}.
In 2013, A. Geim and I. Grigorieva opened new perspectives in this field by proposing the concept of van der Waals heterostructure~\cite{geim-grigorieva_nature2013, novoselov_science2016}, i.e. a compound formed of different monolayers stacked on top of each other, as a way to combine and control the characteristics of several 2D materials on a single system. Since then, such structures have been investigated theoretically and experimentally, and lead to some breakthrough applications and insights~\cite{withers_natmat2015,unucheck_nature2018,cadiz_prx2017}.

Among 2D materials, black phosphorus (BP) exhibits anisotropic mechanical, electronic and optical properties due to its relatively flexible ``accordion" structure. Other peculiarities are a very high hole mobility along the armchair direction~\cite{li_natnanotech2014, liu_acsnano2014,ling-pnas2015,carvalho_natrevmat2016} and, quite uniquely within the 2D family, its fundamental gap remains direct all along the way from the bulk  down to the monolayer. 
As in all 2D materials the layer properties are particularly sensitive to the surroundings, for example the gap width can be modulated on a large range by varying the dielectric properties of the environment~\cite{gaufres_nanolett2019} or the number of stacked layers~\cite{cai_scirep2014,li_natnanotech2017, gaufres_nanolett2019} (from 0.3 in the bulk to more than 2.0 eV in the monolayer). All these characteristics make BP a promising candidate for ultra-flat and flexible opto-electronic technology.
It has been synthesized using various techniques such as mechanical or chemical exfoliation~\cite{li_natnanotech2014,castellanos-gomez_2dmat2014,favron_natmat2015}, or by Chemical Vapor Deposition~\cite{smith_nanotech2016,li_adma2018}. However, at low thickness, its chemical stability is compromised by its high reactivity~\cite{favron_natmat2015}, which implies encapsulation or passivation for any practical application.

Another 2D allotrope of phosphorus is the blue phosphorus (\Pblue), which has been predicted to be stable by theoretical works~\cite{zhu-tomanek_prl2014} and has been produced by molecular-beam epitaxy~\cite{zhang_nanolett2016,xu_prm2017}. It is homostructural to silicene, with a buckled honeycomb lattice. Its electronic properties differ from those of BP, with a larger indirect bandgap, similarly modified by the layer number~\cite{zhu-tomanek_prl2014}.

One way of protecting both the BP and the \Pblue{ }is to cap them with passivation layers. But owing to their sensitivity to the external conditions, it is important to choose capping layers that either preserve target properties, or modify them in the desired way. To this purpose, hexagonal boron nitride (hBN) was proposed as a good candidate because of  its chemical inertness and its insulating properties~\cite{liu-natcom2013, xue-natmat2011} which are predicted to preserve the electronic features of phosphorus layers upon stacking~\cite{cai-zhang-zhang_jpcc2015,doganov_apl2015,hu-hong_applmatinterf2015,constantinescu_nanolett2016,steinkasserer-suhr-paulus_prb2016,zhang-wang-duan_chinphysb2016}.
However, due to lattice mismatch and symmetry differences between the two layers, contact strain cannot be avoided. This is is likely to have an impact on both the structural stability and electronic properties~\cite{birowska_nanot2019}. Since previous studies on P/BN bilayers have taken into account structural relaxation only partially or lack of comparative studies, we think that it is worth carrying out a more systematic investigation.
Complete studies of this kind have been carried out on different systems by Van Troeye and coworkers~\cite{vantroeye_prm2018} on bulk heterostructures of BP on graphene, and by Peng, Wei and Copple~\cite{peng-wei-copple_prb2014} on monolayer BP.

In section I, we present the structures considered in this work. In section II we clarify our methodology and define useful quantities. The structural properties of the BP/hBN and the \Pblue{}/hBN heterostructures are presented and discussed in section III. Section IV is devoted to the discussion on how the electronic properties of the constituent layers are modified by the contact deformations. Conclusions will be presented in section V.

\section{Considered structures}

\begin{figure}
\centering
\includegraphics[height=9cm]{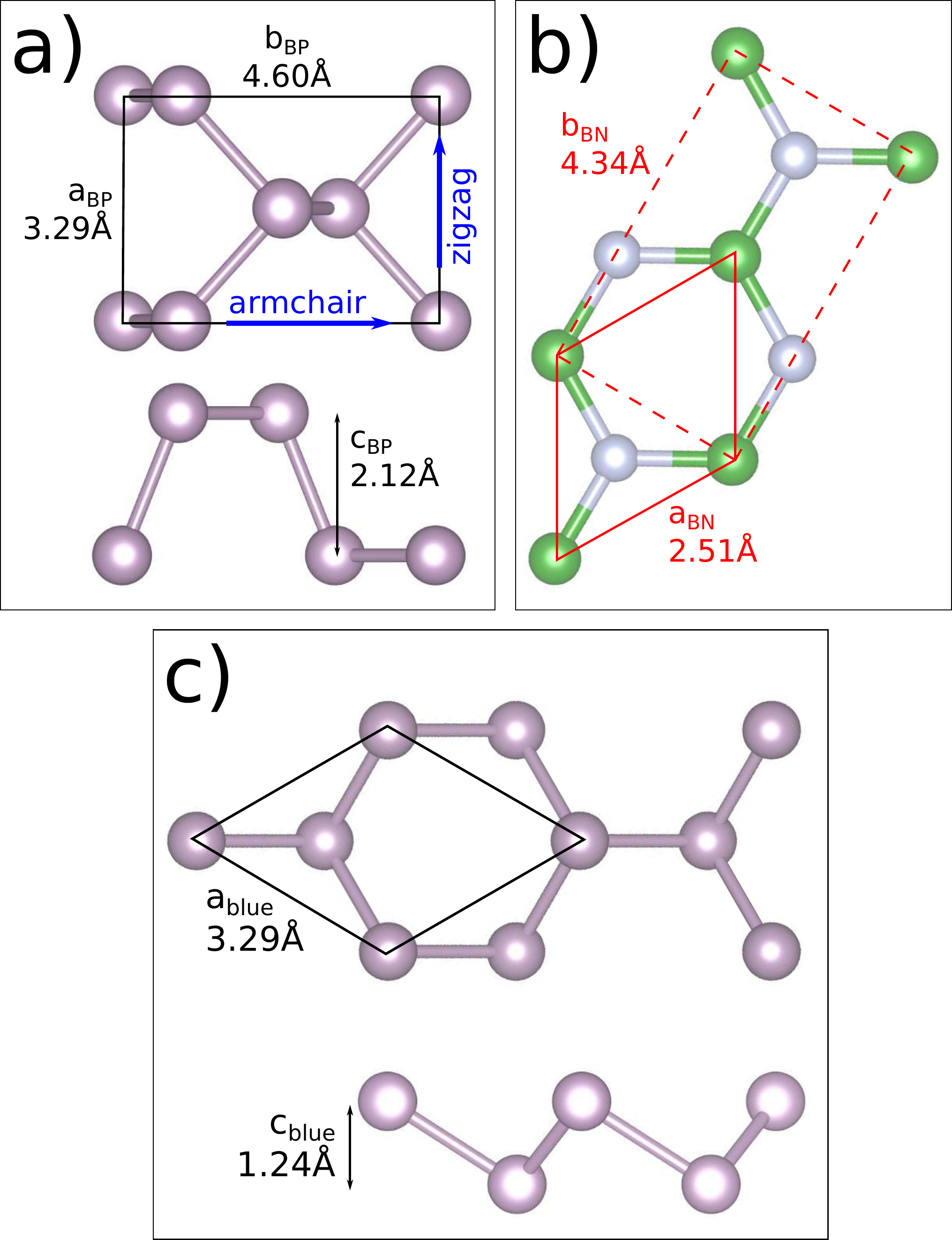}
\caption{a) Top and side view of the relaxed BP monolayer. Blue arrows mark the armchair and the zigzag axis. b) Top view of the hBN monolayer.  The non-unitary cell of hBN is drawn with a dashed line. c) Top and side view of \Pblue{ }monolayer.  All structures are visualized with the VESTA package~\cite{vesta3}.}
\label{fig:f1}
\end{figure}

\begin{figure}
\centering
\includegraphics[width=0.40\textwidth]{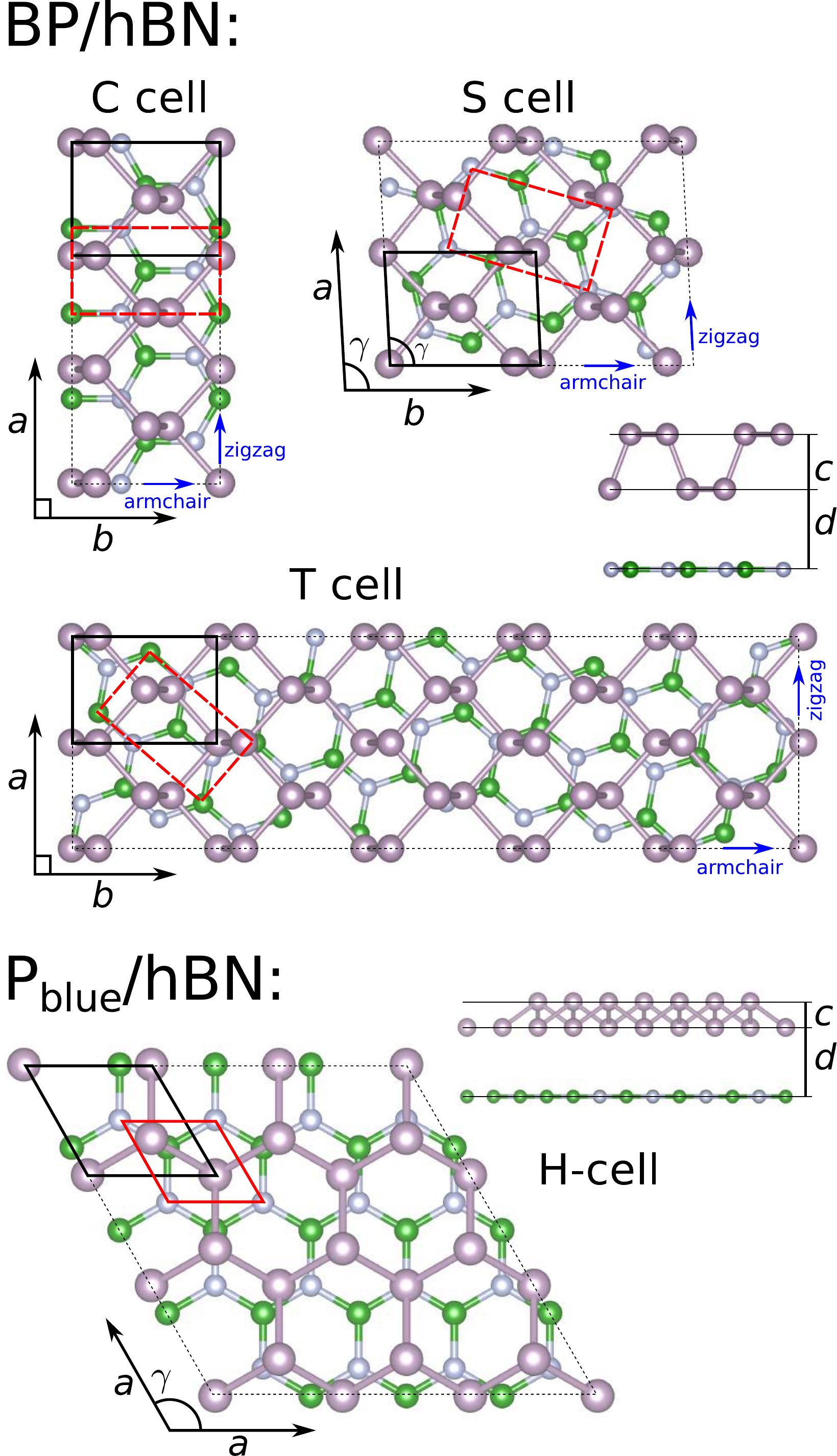}
\caption{Representative cells of the heterobilayers (thin dashed contours). The cells of constituent layers are drawn with thick continuous (unitary) or dashed (non-unitary) lines, with the same color-code of Figure~\ref{fig:f1}.}
\label{fig:f2}
\end{figure}

The building-block monolayers are reported in Figure~\ref{fig:f1}. The structural parameters reported have been obtained after full relaxation (atomic positions and cell parameters) of the unitary cells.
We will refer to the two crystallographic directions of the BP cell using the common ``armchair" and ``zigzag" nomenclature. 
These directions are marked with blue arrows in Figures~\ref{fig:f1} and \ref{fig:f2}.
In constructing the BP/hBN heterobilayers, it is convenient to consider the non-primitive orthorhombic cell of hBN, in which $b_{hBN}=\sqrt{3}a_{hBN}$, which is drawn with a dashed red rectangle in Figure~\ref{fig:f1}b). 

We considered four different heterostructures, three of the BP/hBN type and one of the \Pblue/hBN type.
The corresponding equilibrium structures are visualized in Figure~\ref{fig:f2} and their structural parameters are reported in Table~\ref{tab:cell-bilayers}.
These structures have been chosen to be good compromises between computational feasibility and size.

We start presenting the three BP/hBN systems presented in Fig.~\ref{fig:f2}. 
The C-cell is constructed in such a way that the $b$ axes of the two orthorhombic cells of BP and hBN are equal and parallel (parallel zigzag directions).
It contains 3 BP cells and 8 B-N pairs (4 orthorhombic cells), for a total of 28 atoms. 
At equilibrium, the in-plane $a$ and $b$ parameters are 10.01 and 4.36 \AA{ } respectively, the buckling parameter $c$ of BP is of 2.14 \AA. The found equilibrium distance $d$ between the layers is 3.36 \AA, which falls inside the interval of values reported by other studies done in the same cell~\cite{hu-hong_applmatinterf2015,cai-zhang-zhang_jpcc2015,steinkasserer-suhr-paulus_prb2016}.
In this cell, only tensile and compressive stress are exerted on each layer.
Most of BP/hBN or BP/graphene simulations found in literature~\cite{padilha_prl2015,cai-zhang-zhang_jpcc2015,hu-hong_applmatinterf2015,zhang-wang-duan_chinphysb2016,steinkasserer-suhr-paulus_prb2016,vantroeye_prm2018} are done in the C-cell because it is the smallest supercell accommodating the two layers without excessive deformations.
Still, strain can be quite relevant in this configuration, so it is sensible to look for larger supercells where layers are less strained.

In the S-cell, the BP layer is sheared of $3^\circ$, so that the angle $\gamma$ between the armchair and the zigzag directions is equal to  93$^\circ$ (see Fig.~\ref{fig:f2}).
This deformation allows us to accommodate 4 BP cells on top of 11 B-N pairs, for a total of 38 atoms, without changing significantly the cell parameters of either monolayer with respect to its free-standing structure.
As a result, uniaxial strains are quite low in each sheet, at the price of applying a shearing stress and breaking the orthorhombic symmetry of the BP layer.
The lattice parameters of the S-cell at equilibrium are $a$=6.62~\AA, $b$=9.04~\AA, $c$=2.13~\AA{ }, and $d$=3.43~\AA{}.
This type of cell, already considered in literature~\cite{luo_pbcm2018}, is compared here with other BP/hBN configurations. 
Moreover, shearing the BP layer without deforming the underlying hBN, seems quite a difficult condition to realize in actual experiments.

In the T-cell, strain is minimized through a twist angle of $19.14^\circ$ between the orthorhombic cells of the two layers. 
In this configuration, it is possible to fit 10 unitary cells of BP on top of 28 B-N pairs, for a total of 96 atoms.
The structural parameters of the supercell at equilibrium are 
$a$=6.62~\AA, $b$=22.95~\AA, $c=$2.11~\AA{}, and $d=$3.43~\AA.
As in the C-cell, no other deforming force is exerted on the layers of the T-cell apart from tensile and compressive stress.
The same twist angle has been investigated by Van Troeye and collaborators~\cite{vantroeye_prm2018} who scrutinized also other twisted configurations, however the systems they studied are BP/graphene bulk heterostructures, so quite different from ours in particular because of the absence of polar atomic bonding in graphene.
Finally, Costantinescu and coworkers~\cite{constantinescu_nanolett2016} have investigated similar systems (hBN/BP/hBN and BP/hBN/BP) with a twist angle of about 40$^\circ$ between the layers, but the study lacks comparison with other structures.  

The only cell type considered in the \Pblue{}/hBN bilayer is reported in Figure~\ref{fig:f2} and is called H-cell because of its in-plane hexagonal symmetry.
It is constructed by fitting 9 unit cells of \Pblue{ }on top of 16 unit cells of hBN, for a total of 50 atoms.
In the resulting configuration, only biaxial tensile or compressive stresses are applied on the layers.
Its equilibrium lattice parameters are: $a$=9.99 \AA,  $c$=1.23 \AA{ }, and $d$=3.37 \AA.

\begin{table}[t]
\centering
    \begin{tabular}{c|c c c c}
        \hline \hline
         Cell type &  $a$  & $b$   & $c$  & $d$\\
        \hline
        C-cell     & 10.01 &  4.36 & 2.14 & 3.36\\
        S-cell     &  6.62 &  9.04 & 2.13 & 3.43\\
        T-cell     &  6.62$^\dagger$ & 22.95$^\dagger$ & 2.11 & 3.43\\
        \hline
        H-cell     & \multicolumn{2}{|c}{9.99} & 1.23 & 3.37\\
        \hline \hline
    \end{tabular}
    \caption{Cell parameters of the equilibrium P/BN bilayers.\\$^\dagger$ The equilibrium parameters of the T-cell are derived from a fit, not from a full relaxation (see text).}
    \label{tab:cell-bilayers}
\end{table}

\begin{figure}[b]
    \centering
    \includegraphics[width=0.41\textwidth]{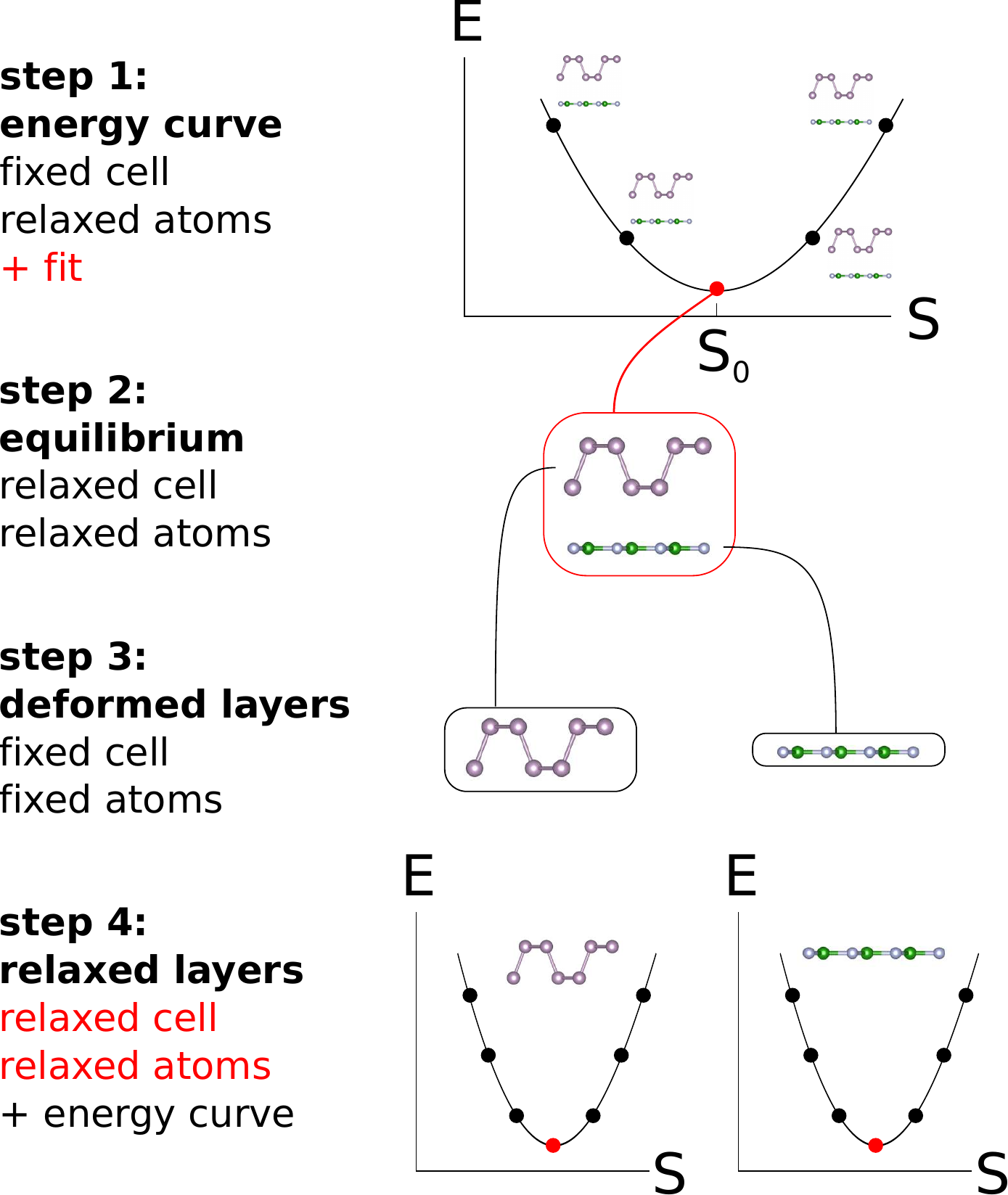}
    \caption{Cartoon of the relaxation procedure adopted in this work}
    \label{fig:relaxation}
\end{figure}

\section{Computational details and methods}

All first-principle calculations in this paper are done within density functional theory (DFT) with the ABINIT simulation package~\cite{gonze_cms2002}. The Kohn-Sham states are represented with a plane-wave basis with a cutoff energy of 60 Ha.
Norm-conserving pseudopotentials~\cite{hamann_prb2013} are used, including 3 electrons for B and 5 electrons for N and P.
The exchange-correlation potential is approximated using the GGA-PBE functional~\cite{perdew_prl1996}.
In addition, van der Waals forces are taken into account with the vdW-DFT-D3 scheme introduced by Grimme~\emph{et al.}~\cite{grimme_jpc2010}.
The in-plane k-point sampling used in the density calculations has one point along $z$ and 5x3 in the plane of the C-cell, 4x4 in the S-cell, 3x5 in the T-cell and 5x5 in the H-cell.
To avoid artifacts due to interacting periodic replicas along the $z$ axis, a vacuum of 16 \AA{ }is included in all supercells.

The heterobilayers considered in this study can be divided into four groups: the C-cell, S-Cell, T-cell and H-cell.
In order to analyse the energetic costs and gains to first deform and then stack the layers, we employed the structural optimization procedure sketched in Figure~\ref{fig:relaxation}.

\begin{enumerate}
    \item The first step consists of varying by hand the size of the cell while relaxing only the atomic positions.
    In doing this, we took care of preserving the hexagonal symmetry of the hBN layer, which fixes a $b/a$ ratio in each cell.
    This procedure allowed us to draw stability curves of the bilayers, i.e. total energies as functions of the surface $S$.
    Successively we fitted these curves with the quadratic function
    \begin{equation}
        E(S) = E(S_0) + \frac{B}{4S_0} \left(S-S_0\right)^2 \,
        \label{eq:energy-curve}
    \end{equation}
    where $E(S)$ is the total energy at surface $S$, $S_0$ the surface which minimizes the energy fit, and $B=S_0 d^2E/dS^2$ the uniaxial or biaxial 2D compressibility depending on the nature of the strain. We prefer expressing this quantity in meV/\AA$^2$ instead of GPa, as commonly done because that would imply a normalisation on the thickness of the layer, which is a quite arbitrary concept we want to avoid.
    The minimum of the stability curve $E(S_0)$ constitutes the starting point for the next step.
    
    In the case of the C-cell, 
    this procedure has been initialized with eight inequivalent configurations, each corresponding to one way of stacking the four atoms of a unitary cell of BP on top of the two atomic species of hBN. 
    For the other cells, 
    the large size of the calculation hampered the exploration of several initial configurations.
    
    \item Once $S_0$ is extracted from the fit, we allow for a full structural relaxation (atomic positions and cell parameters), with the intent to account for possible deviations from the fitting functions. 
    At this step, the hexagonal symmetry of hBN may break due to internal stresses.
    The resulting structures, the parameters of which are reported in Table~\ref{tab:cell-bilayers}, are our best estimate of the equilibrium configurations of the heterobilayers.
    Note that in the T-cell a full relaxation has not been possible owing to its large dimensions. Instead, cell parameters have been derived from the fitted $S_0$ and the hexagonal symmetry of hBN ($b/a = 3.464$) and then the relaxation has been done only for the atomic positions.
    
    \item Successively, from the equilibrium heterostructures, we remove all the P atoms and we compute the total energy of the hBN which is left without performing any relaxation. So the hBN layer is isolated but its atoms occupy the same position as in the equilibrium heterostructure (deformed monolayer).
    Then we do the same by keeping only the phosphorus atoms and removing the hBN.
    In this way we can access the total energy of a hypothetical stage where each building-block is isolated but experiences the same deformations as in the bilayer.
    
    \item Finally, fully relaxed calculations of the monolayers in their unitary cells gave us access to the total energy of the free-standing pristine materials (relaxed monolayers).
    
\end{enumerate}
We use the FIRE~\cite{fire} algorithm for the first optimization steps, and successively the BFGS~\cite{bfgs} as they are implemented in ABINIT.
The stopping condition for all relaxation calculations is that the change of the forces exerted on every atom must be lower than $10^{-5}$ Ha/Bohr along all Cartesian directions.

Within the heterobilayer, the strain percentage on the generic lattice parameter $p$ of the material M (M=BP, \Pblue{ }or hBN) is defined 
\begin{equation}
    \eta(p_M) = 100 \left( p_M - p^{\rm REF}_M \right)/p^{\rm REF}_M \,,
    \label{eq:strain}
\end{equation}
where $p^{\rm REF}_M$ is the corresponding cell parameter of the relaxed monolayer M.

\section{Structural properties}

In this section, we first investigate the influence of induced deformations on the total energy of the three isolated monolayers, then we pass to the discussion of the heterobilayers.

\subsection{Isolated monolayers}

\begin{figure}
    \centering
    \includegraphics[width=0.48\textwidth]{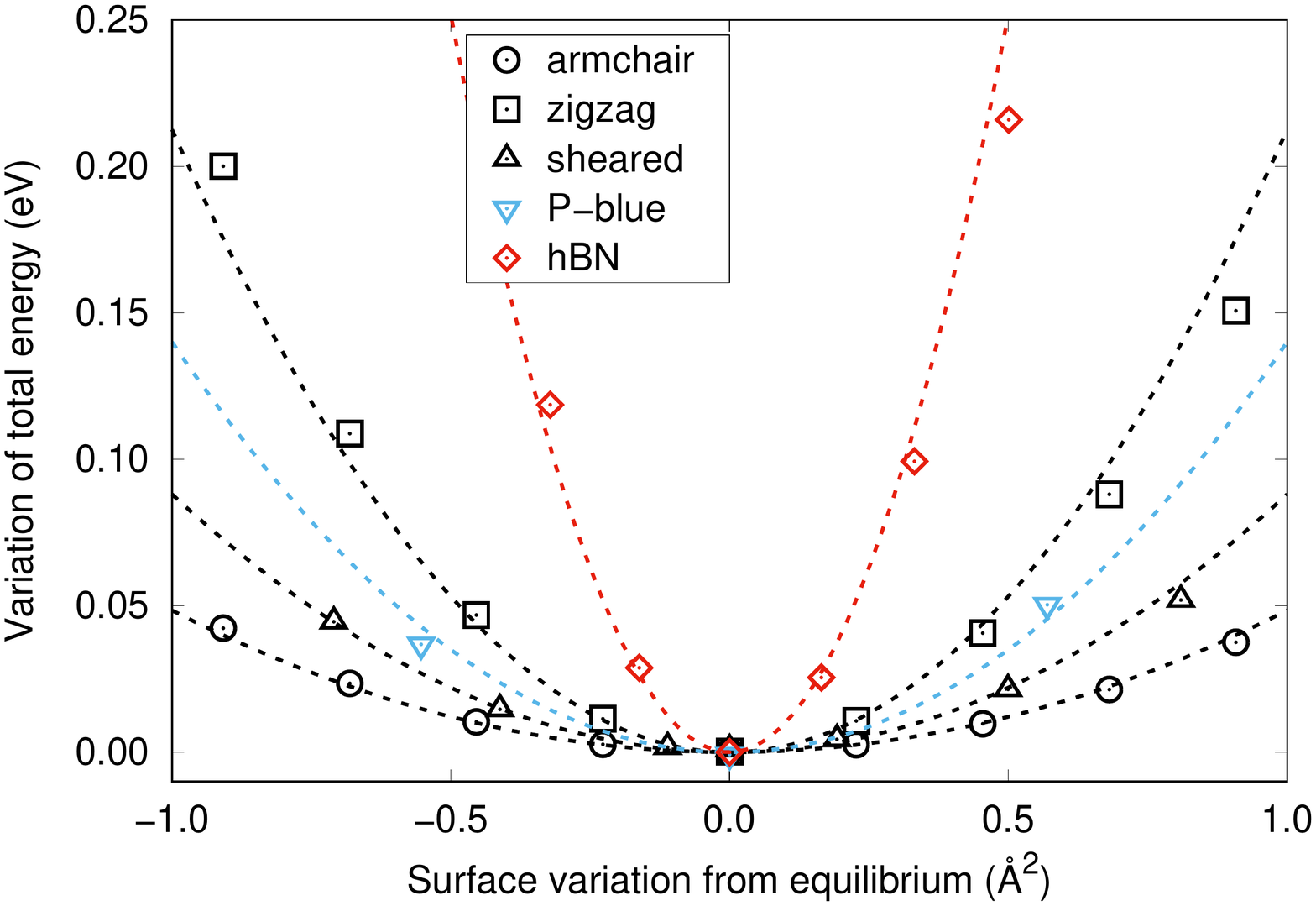}
    \caption{Stability curve of the monolayers. Black Circles: BP strained along the armchair axis; Black squares: BP strained along the zigzag axis;  Black upward-pointing triangles: sheared BP biaxially strained; 
    Blue downward-pointing triangles: \Pblue{} biaxially strained; Red diamonds: hBN biaxially strained. Dashed colored lines: fit of ~\ref{eq:energy-curve}.}
    \label{fig:ecurve_all-monolayers}
\end{figure}

The equilibrium structure of the monolayers, reported in Figure~\ref{fig:f1} have been obtained using a full-relaxation procedure (step 4).
In addition, starting from each equilibrium structure, we changed the cell parameters and let the atomic positions relax in order to compute total energy variations as functions of the resulting strain.
This allowed us to draw the stability curves reported in Figures~\ref{fig:ecurve_all-monolayers}, from which we derived the 2D compressibility $B$ of the monolayers by fitting equation~\ref{eq:energy-curve} (see Table~\ref{tab:monolayers} and Figure~\ref{fig:ecurve_all-monolayers}).
Our stability curves of BP and hBN are in good agreement with other published results~\cite{peng-ji-de_compmatsc2012,peng-wei-copple_prb2014}.

\begin{table}[b]
    \centering
    \begin{tabular}{r | c c }
        \hline \hline 
          & $S_0$ & $B$ \\ 
         \hline
        hBN             &  5.45 & 21.93 \\
        BP (armchair)   & 15.14 &  2.93 \\
        BP (zigzag)     & 15.14 & 12.87 \\
        BP (sheared)    & 15.23 &  5.36 \\
        \Pblue             &  9.36 &  5.24 \\
        \hline \hline
    \end{tabular}
    \caption{Equilibrium surface $S_0$ (in \AA$^2$) and 2D compressibility $B$ (in eV/\AA$^2$) of the monolayers resulting from the fit of expression~\ref{eq:energy-curve}.}
    \label{tab:monolayers}
\end{table}

\begin{table*}
    \centering
    \begin{tabular}{r | c | c c | c c c}
    \hline \hline
     & $B$ &\multicolumn{2}{c|}{Strain hBN (\%)} & \multicolumn{3}{c}{Strain phosphorus (\%)}\\
    Cell type      & meV/\AA$^2$ & armch. & zigzag & armch. & zigzag & buckling \\
    \hline
    C-cell         & 25.36   &  0.29  & -0.16  & -5.32 & 1.46 & 0.77\\
    S-cell      &    25.22   &  0.02$^*$  & -0.15$^*$  & -1.83 & 0.82 & 0.50\\
    T-cell         & 26.73   & -0.14$^*$  & -0.14$^*$  & -0.20 & 0.71 & 0.16\\
    \hline
    H-cell  &        27.94 & \multicolumn{2}{c|}{-0.40} &  \multicolumn{2}{c}{1.28} & -2.18\\
    \hline \hline
    \end{tabular}
    \caption{2D compressibility and strain distribution on the constituent layers.  In the H-cell biaxial strain is along both $a$ directions.\\$^*$ Because of the misalignment in the S-cell and the T-cell, the strain distribution in hBN is always given with respect to the armchair and zigzag directions of the BP layer (cfr Figure~\ref{fig:f2}). Note that they are not perpendicular in the S-cell.}
    \label{tab:equilibrium_BPhBN}
\end{table*}

In the case of hBN (red diamonds) and \Pblue{ }(blue downward-pointing triangles) we have applied a biaxial stress to preserve their hexagonal symmetry, while in the case of the orthorhombic BP layer, we have varied the $a$ (zigzag) and the $b$ (armchair) parameters separately, which corresponds to applying uniaxial stress along either direction.
In this study, we are not interested in biaxially strained BP because, once inside the bilayer, its deformations are anisotropic. For a study on the effects of biaxial strain on BP, we refer to the work by \c{C}ak\i{}r and coworkers~\cite{cakir-sahin-peeters_prb2014}.
Finally, we have looked at the stability curve of the sheared BP monolayer (black upward-pointing triangles), where we decided to apply stress in such a way to keep a shear angle of $3^\circ$.

According to our analysis, the hBN is clearly the stiffest layer of all owing to its flat hexagonal structure which hinders any deformation. As expected~\cite{peng-wei-copple_prb2014,vantroeye_prm2018}, the BP structure is found to be much softer along the armchair direction than along the zigzag one.
The \Pblue{ }layer is also quite soft because the in-plane stress can be partially transferred to the buckling parameter $h$, which passes from 1.29 \AA{ } when $a$ is compressed of 6\% to 1.20 \AA{ } upon a dilatation of the same amount.

As a result, in the P/BN heterostructures we expect strain to be larger on the phosphorus sheet (BP or \Pblue{ }) which will deform conveniently to adapt to the much more rigid hBN layer.

\subsection{BP/hBN bilayer}

In this section we discuss the effects of strain on the three BP/hBN cells.
Results are summarised in Table~\ref{tab:equilibrium_BPhBN}.

\subsubsection{Cell type C}
As explained already in the methodological section, in the case of the C-cell type, we have been able to carry out the relaxation procedure (step 1) starting from eight different configurations.
Half of them correspond to the four possibilities to place the inequivalent P atoms of BP on top of a N atom placed at the origin (P/N configurations).
In the other four configurations, the same P atoms are on top of a B atom placed at the origin (P/B configurations).
Note that other configurations were possible, for example we could have put the P-P bonding on top of B or N, but configurations alike have been predicted to be less stable~\cite{cai-zhang-zhang_jpcc2015}.
Moreover, they can be obtained from those we considered through a rigid translation in the armchair direction of the BP, but such a modification has been demonstrated to be energetically expensive~\cite{padilha_prl2015,hu-hong_applmatinterf2015,zhang-wang-duan_chinphysb2016}.

\begin{figure}[b]
    \centering
    \includegraphics[width=0.48\textwidth]{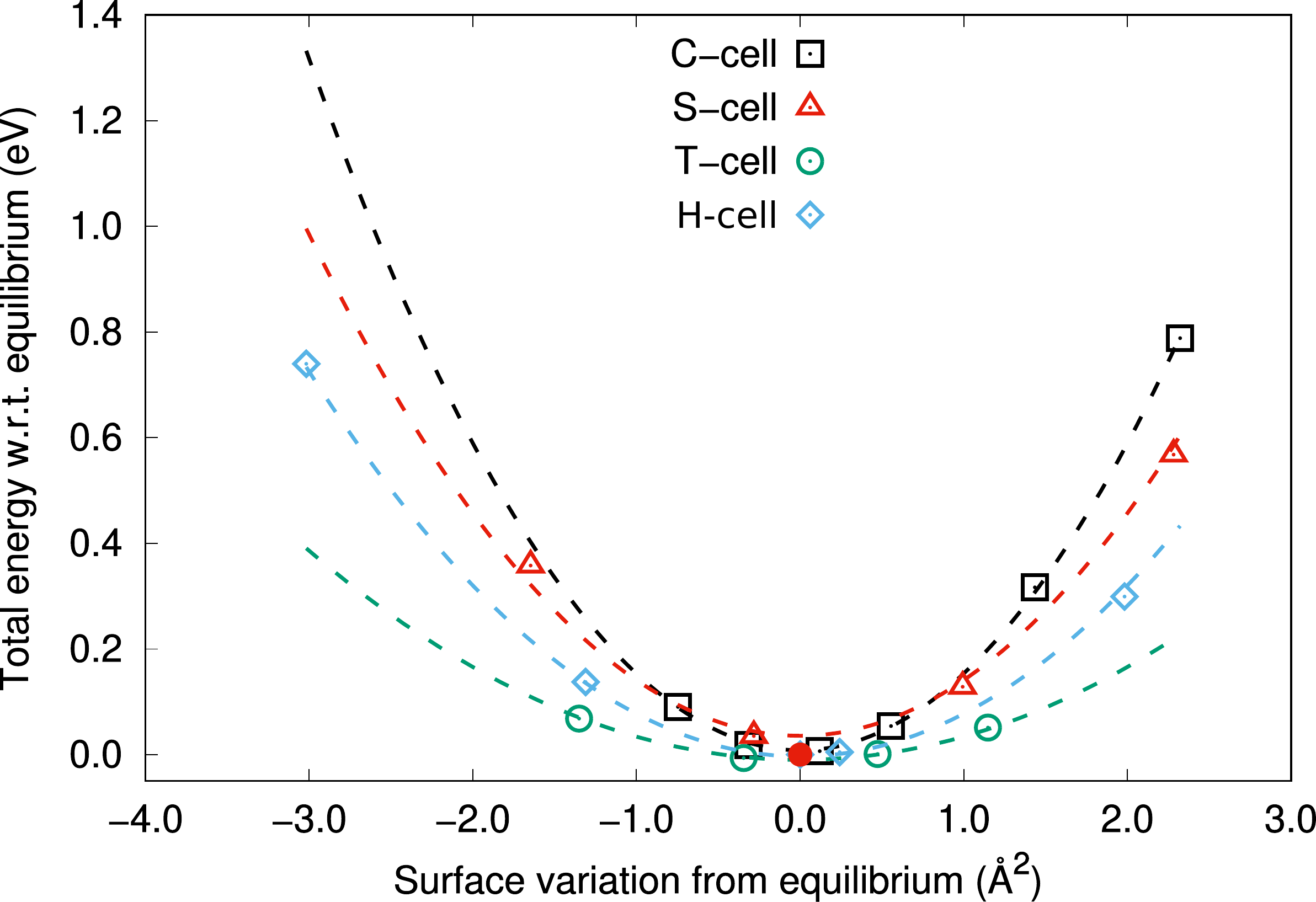}
    \caption{Phosphorus/hBN cells: Total energy as a function of the deviation from the equilibrium surface $\Delta S=S-S_0$. Black squares: C-cell; Red triangles: S-cell; Green circles: T-cell, Blue diamonds: H-cell. Dashed colored lines: fit of~\ref{eq:energy-curve}. The red bullet at the origin marks the equilibrium structure of each system.}
    \label{fig:ecurve_all-BPhBN}
\end{figure}

The resulting eight stability curves (not shown) have the same shape and all relax to the same cell-parameter with negligible total energy differences (order of 0.1 meV).
However, it is possible to identify some structural selection mechanism since, at the end of the relaxation procedure, all the P/N cells have evolved into P/B ones.
This tendency is rationalized observing that in the B-N ionic bonding, trivalent B atoms have a positive charge, whereas pentavalent N atoms are negatively charged. As P is also pentavalent, and hence attracted by B atoms. 
This repulsion is obviously not present in BP/graphene~\cite{vantroeye_prm2018} and is likely related to the polar nature of the B-N bonding.

All further analysis on the characteristics of the C-cell have been carried out on the structure with the lowest total energy, chosen as representative in virtue of the very tiny energy and structural differences among them.
In Figure~\ref{fig:ecurve_all-BPhBN}, black squares mark the computed total energy as a function of the surface variation $\Delta S$ of this representative C-cell. In dashed black, the fit of expression~\ref{eq:energy-curve} allowed us to extract the biaxial 2D compressibility $B$ of the bilayer, as explained in Section II (step 1).

In order to form the heterostructure, both layers deform.
In particular, the BP layer contracts along the armchair axis (-5.32\%) and elongates along the zigzag axis (+1.44\%) and the buckling (+0.77\%).
Similarly, the hBN layer undergoes an anisotropic deformation but with opposite sign, with a weak compression (-0.16\%) along the zigzag axis aligned to the BP zigzag one, and a weak dilatation (0.29\%) of the armchair axis perpendicular to the former. 
As expected, the preferential direction to release the stress is by far the relatively soft armchair axis of BP, while the hBN layer acts basically as a rigid support.

\subsubsection{Cell type S}

Data from the S-cell are reported in red triangles in Figure~\ref{fig:ecurve_all-BPhBN}.
The 2D biaxial compressibility of the S-cell is almost the same as that of the C-cell.



In this configuration, the strain is actually evaluated along directions that are not Cartesian, the shear angle between the BP (reference) armchair and zigzag direction being 3$^\circ$.
Moreover in the hBN layer these two directions have no special interest (neither is the zigzag or armchair axis of hBN).
So the analysis of the strain is somewhat more involved and results should be interpreted as qualitative indications.
What can be said, however, is that also in this case most of the stress applies along the armchair axis of the BP layer which compresses to accommodate to the underlying hBN sheet, even though strain is much lower than in the C-cell, which is the reason it has been studied in literature~\cite{luo_pbcm2018}. 
Similarly to the previous case, the hBN layer plays the role essentially of a rigid substrate, undergoing only a very weak and anisotropic deformation with a different sign with respect to that of the BN layer. 

\subsubsection{Cell type T}

Because of the large dimension of this cell-type, once we extracted the equilibrium surface $S_0$ from the fit,
we could not make a full relaxation to permit anisotropic deformations of the hBN layer.
So we constructed a cell of surface $S_0$ with fixed ratio $b/a=3.464$, which keeps the hexagonal symmetry of the hBN layer (cfr. Table~\ref{tab:equilibrium_BPhBN}), and we relaxed only the atomic positions.
From this equilibrium position we changed the cell parameters to construct the energy curve, reported in green circles in Figure~\ref{fig:ecurve_all-BPhBN}.
Thanks to the twist angle between the BP and the hBN, this structure minimizes the deformations on the BP layer while keeping the deformation of hBN quite low (hBN: -0.14\% biaxial; BP: -0.2\% on armchair, 0.71\% on zigzag and 0.16\% on buckling). In addition it is found to be the most rigid of all bilayers.

In the work of Van Troeye and collaborators~\cite{vantroeye_prm2018} on BP/graphene heterostructures, the BP layer always deforms by contracting along the armchair axis, so the authors suggest that this is a general behaviour of BP under stress.
This is indeed what we observed in the S-cell and the C-cell.
However, in the T-cell, most of the deformation undergone by the BP layer is a dilatation of the zigzag axis, despite its higher stiffness, whereas the armchair axis contracts only by 0.2\%.
This result can be ascribed to two differences with respect to the work of Van Troeye.
First, our BN layer is less homogeneous than the graphene layer because of the polar nature of the B-N bonding, which acts on a short-range scale, probably through the same P/N repulsion we observed while relaxing the C-cells.
We expect this characteristic to cause an angle-dependent deformation behavior of the deposited BP layer.
Second, Van Troeye's systems are bulk heterostructures, not isolated bilayers. This can also influence the atomic arrangement through long-range interactions, even though we estimate these long-range contributions to be less important than the short-range ones.
Unfortunately, our computational resources could not allow us to investigate further these hypotheses.

\subsubsection{H-cell}
The data relative to the H-cell are reported in Table~\ref{tab:equilibrium_BPhBN} and as blue diamonds in Figure~\ref{fig:ecurve_all-BPhBN}. 

In relaxing this structure we applied biaxial stress on the bilayer supercellmaking the hypothesis that the hexagonal symmetry is kept in deformation because it is common to both layers.
The results confirm once more that the hBN has the tendency to contract slightly (-0.4\%) while the P layer undergoes much larger deformations to adapt to the underlying sheet.
In this case, the biaxial in-plane dilatation of the \Pblue{ }layer (+1.28\%) is accompanied by a sizable reduction of the buckling of 2.18\%.

\subsubsection{Stability and energy costs}

\begin{figure}
    \centering
    \includegraphics[width=0.48\textwidth]{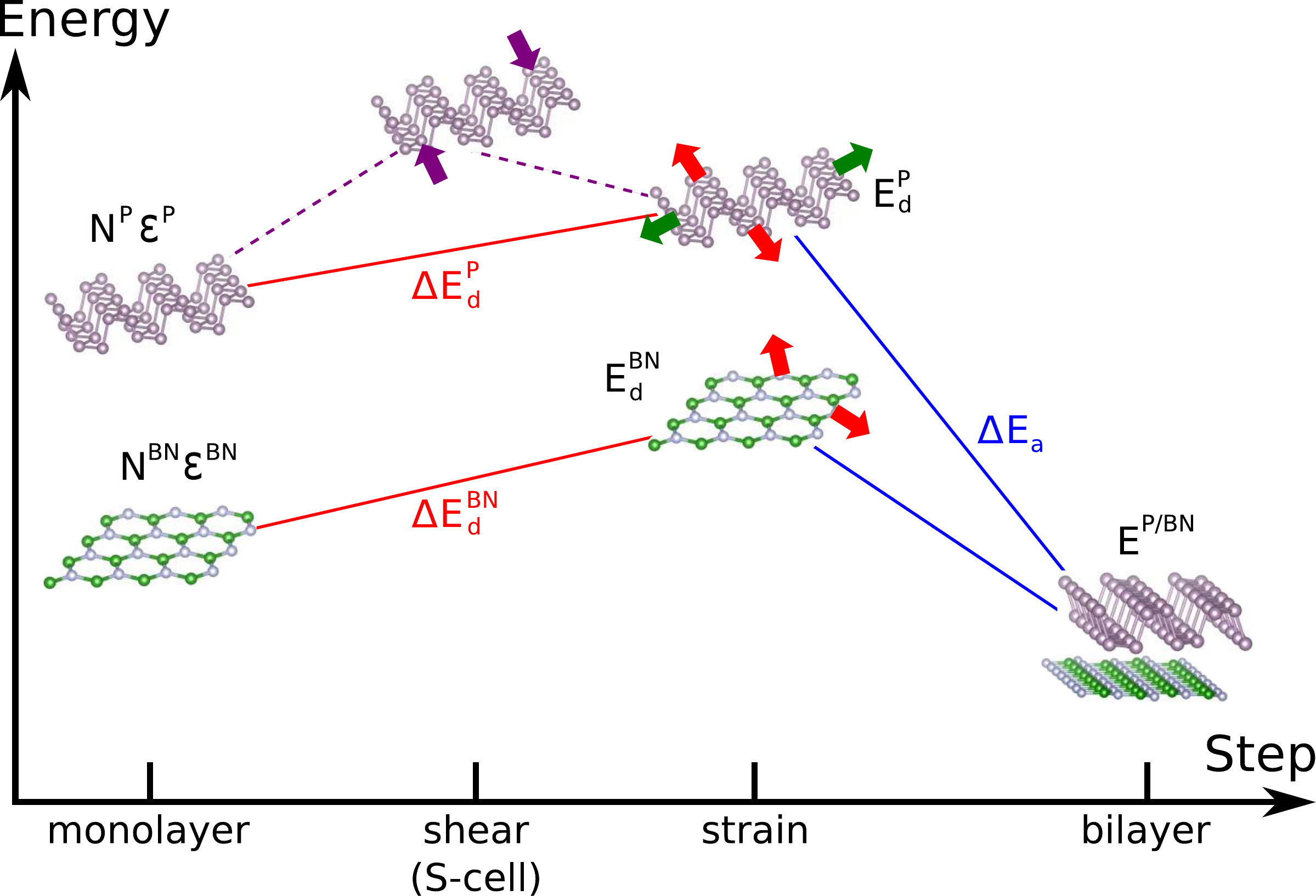}
    \caption{Schematic representation of the steps from isolated monolayers to the heterostructure. Colored arrows represent the deformation applied.}
    \label{fig:energy_costs}
\end{figure}

Our relaxation procedure allowed us to quantify the stability of the bilayers, while distinguishing how much energy is employed to deform the layers and how much is gained in the adhesion, as depicted schematically in Figure~\ref{fig:energy_costs}.

In order to compare the stability of the different bilayers, we introduce the formation enthalpy per unit surface
\begin{equation}
    \Delta H =\frac{E^{P/BN} - N^P \mathcal{E}^P  - N^{BN}\mathcal{E}^{BN}}{2S^{P/BN}} = \frac{\Delta E}{2S^{P/BN}}\;,
     \label{eq:enthalpy}
\end{equation}
where $N^M$ is the total number of atoms belonging to the P or the BN layer in the unitary cell of the bilayer, $\mathcal{E}^M$ is the total energy per atom of the isolated monolayer of the material M, $E^{P/BN}$ is the total energy of the unitary cell of the bilayer and $S^{P/BN}$ its surface.
The formation enthalpy is negative if the structures is stable and the lower it is, the more stable is the bilayer.
It is simply related to the binding energy per P atom $E_b=\Delta E/N^P$ used in similar studies~\cite{padilha_prl2015,hu-hong_applmatinterf2015,cai-zhang-zhang_jpcc2015,steinkasserer-suhr-paulus_prb2016,zhang-wang-duan_chinphysb2016} and to the cohesion energy per atom $E_{coh}=\Delta E/(N_P+N_{BN})$ employed by Van Troeye \emph{et al.}~\cite{vantroeye_prm2018}.

The number of atoms, the bilayer surface and the formation enthalpy are reported in the top part of Table~\ref{tab:energy-costs}.
All bilayers are predicted to be stable.
Among the BP/hBN cells, the S-cell exhibits clearly the weakest cohesion. 
On the other hand, the H-cell has the lowest formation enthalpy among all considered bilayers, comparable to that of the S-cell, despite the same symmetry of the two layers, although this result may result from the specific choice of the stacking.
The difference between the enthalpy of the C-cell and the T-cell is very low so we are not in the condition of claiming that one is particularly more stable than the other. 
This result marks a difference from the conclusions of drawn for hBN/graphene bulk heterostructures~\cite{vantroeye_prm2018}, where the C-cell was clearly favoured.
As discussed earlier, this different conclusion can come from two factors:  one connected to the weak polar nature of the B-N bonding and to the resulting P-N repulsion we observed, and the other to the different geometries considered in the two studies (isolated bilayers here, bulk heterostructures in~\cite{vantroeye_prm2018}).

For comparison, we checked that our binding energy of the C-cell ($E_b=-62.2$ meV/P) falls in the interval of values reported in literature~\cite{cai-zhang-zhang_jpcc2015,hu-hong_applmatinterf2015,zhang-wang-duan_chinphysb2016,steinkasserer-suhr-paulus_prb2016}.
These range from a maximum of -84 meV/P~\cite{zhang-wang-duan_chinphysb2016} to a minimum of -49 meV/P~\cite{hu-hong_applmatinterf2015}$^,$\footnote{In the work of Hu and Hong~\cite{hu-hong_applmatinterf2015}, the binding energy is actually given in meV per atom of BN. The value we report in the main text is a conversion of the published one.}.

We analysed in more details the energy cost of the evolution from the monolayers to the bilayer by splitting the formation enthalpy into three contributions $\Delta H = \Delta E^{P}_{d} +\Delta E^{BN}_{d} + \Delta E_{a}$.
To this end, we introduce the deformation energy per unit surface of the sheet (M=P or BN)
\begin{equation}
    \Delta E_d^{M} = \left(E^M_d - N^M \mathcal{E}^M\right)/2S^{P/BN}\,,
    \label{eq:deformation-enthalpy}
\end{equation}
defined as the energy difference between the isolated monolayer deformed as in the bilayer ($E^M_d$) and the fully relaxed monolayer, divided by twice the surface of the bilayer.
This quantity expresses the energy paid per unit surface to deform each monolayer to the same configuration it will have inside the bilayer.
The adhesion energy
\begin{equation}
    \Delta E_a = \left(E^{P/BN}- E^{P}_d - E^{BN}_d\right)/2S^{P/BN}
    \label{eq:adhesion-enthalpy}
\end{equation}
is defined as the energy cost per unit surface to stack the two monolayers once they have the appropriate atomic configuration.

\begin{table}[b]
    \centering
    \begin{tabular}{r|c c c | c}
    \hline \hline
                            &  C-cell & S-cell   & T-cell  & H-cell\\
    \hline
$N^{P}$                     & 12      & 16       & 40      & 18   \\ 
$N^{BN}$                    & 16      & 22       &   56    &  32   \\
$S_{P/BN}$                  & 43.62   &  59.82   & 152.02  & 86.36 \\ 
$\Delta H$                  & -8.55   & -7.83    & -8.69   & -7.78  \\
\hline    
$\Delta E^{BN}_{d}$         & 0.13    & 0.19     &  0.02   & 0.38 \\ 
$\Delta E^{P}_{d}$          & 1.79    & 1.56     &  0.69   & 0.41 \\ 
$\Delta E_{a}$              &-10.47   & -9.58    & -9.40   & -8.57\\ 
    \hline \hline
    \end{tabular}
    \caption{Formation enthalpy, deformation energy and adhesion energy of the considered bilayers. For the symbols, refer to expressions~\ref{eq:enthalpy}-\ref{eq:adhesion-enthalpy} and to Figure~\ref{fig:energy_costs}. Units are in \AA$^2$ and meV/\AA$^2$.}
    \label{tab:energy-costs}
\end{table}

The comparison between $\Delta E_a $ and $\Delta E_d^M$ tells us the importance of taking into account deformations.
Among the BP/hBN cells, the configuration where deformations cost the most is the widely studied C-cell, while that where they cost the least is the T-cell. In absolute terms however, it is in the H-cell where the deformation energy is minimized.
More details can be understood if looking at the strain distributions reported in Table~\ref{tab:equilibrium_BPhBN}. In all BP/hBN structures, the deformation energy concentrates essentially on the softer BP layer, but not with the same effects. Indeed in the C-cell, with the largest deformation energy, such a cost is explained by the amplitude of the compression of the armchair axis (-5.32\%).
In the S-cell the amount of axial strain is substantially lower, so most of the deformation energy is used to shear the structure. The result is that, despite the lower axial strain, the deformation energy of the S-cell is comparable to that of the C-cell.
Finally, in the T-cell, the quite low deformation energy (less than half of that of the S-cell) is spent mostly to elongate weakly the relatively rigid zigzag axis. The low price of its deformations is the reason for the higher stability of the T-cell.

In fact, the energy hierarchy is actually inverted once the two layers are stacked.
The lowest $\Delta E_a$ is the one of the C-cell, while the T-cell has the highest one among the BP/hBN family. Despite this, the respective deformation energies are such that the T-cell turns out to be the most stable structure, with a $\Delta H$ comparable to that of the C-cell. 
The S-cell instead, having already the worst $\Delta E_a$ results in being by far the less stable among the BP/hBN bilayers.

Concerning the H-cell, it exhibits the worse $\Delta H$ despite the quite low deformation energies, comparable to those of the T-cell. Similarly to the S-cell scenario, this has to be ascribed to the bad $\Delta E_a$, more than to the deformation term since this is as low as in the T-cell.
Quite interestingly, the higher rigidity of the \Pblue{ }layer forces the hBN to deform more than in the BP/hBN heterostructures.
However, in the case of the \Pblue/hBN structures, we stress that a single configuration is not enough to generalise this result and it could be that other \Pblue/hBN bilayers exhibit higher stability.

\section{Electronic properties}

In previous works on BP/hBN heterostructures~\cite{cai-zhang-zhang_jpcc2015,hu-hong_applmatinterf2015,steinkasserer-suhr-paulus_prb2016,zhang-wang-duan_chinphysb2016}, the bilayers were always in the C-cell configuration, which by itself imposes a lot of stress onto the BP layer, even when relaxed, as we have shown in the previous section.
Shrinking the hBN layer instead of the BP one has also been considered~\cite{zhang-wang-duan_chinphysb2016}, but we have shown that this is not a realistic account of the strain conditions in this system.
These works all agree on the fact that hBN is a promising candidate for protecting BP while preserving its peculiar electronic properties, in particular its high hole mobility in the armchair direction~\cite{liu_acsnano2014}.
However, owing to the extreme sensitivity of BP properties upon deformation~\cite{peng-wei-copple_prb2014}, we think it is worth reconsidering this statement on the light of the results we presented above on the distribution of strain.

To this end, the electronic properties of the bilayers are discussed on the basis of the strain distribution of their constituent sheets.
In Table~\ref{tab:hs-gap} we report the PBE energy gap of the four heterostructures. 
It is well known that DFT systematically underestimates the band gap of semiconductors and its value strongly depends on the choice of the exchange-correlation potential.
However differences are quite reliable within the same approximation.

\begin{table*}
    \centering
    \begin{tabular}{c|c c c |c c c c}
    \hline \hline
                & \multicolumn{3}{c|}{relaxed monolayers} & \multicolumn{4}{c}{P/BN bilayers}\\
                   & hBN & BP & \Pblue & C-cell   & S-cell    & T-cell  & H-cell \\
    \hline
        $E_g$ (eV) & 4.71 & 0.81 & 1.93   & 0.83     & 0.86      & 0.91    &  1.81 \\ 
    \hline \hline
    \end{tabular}
    \caption{PBE energy gap in the isolated monolayers and the four bilayers.}
    \label{tab:hs-gap}
\end{table*}

\subsection{The BP/hBN bilayers}

\begin{figure}
    \centering
    \includegraphics[width=0.48\textwidth]{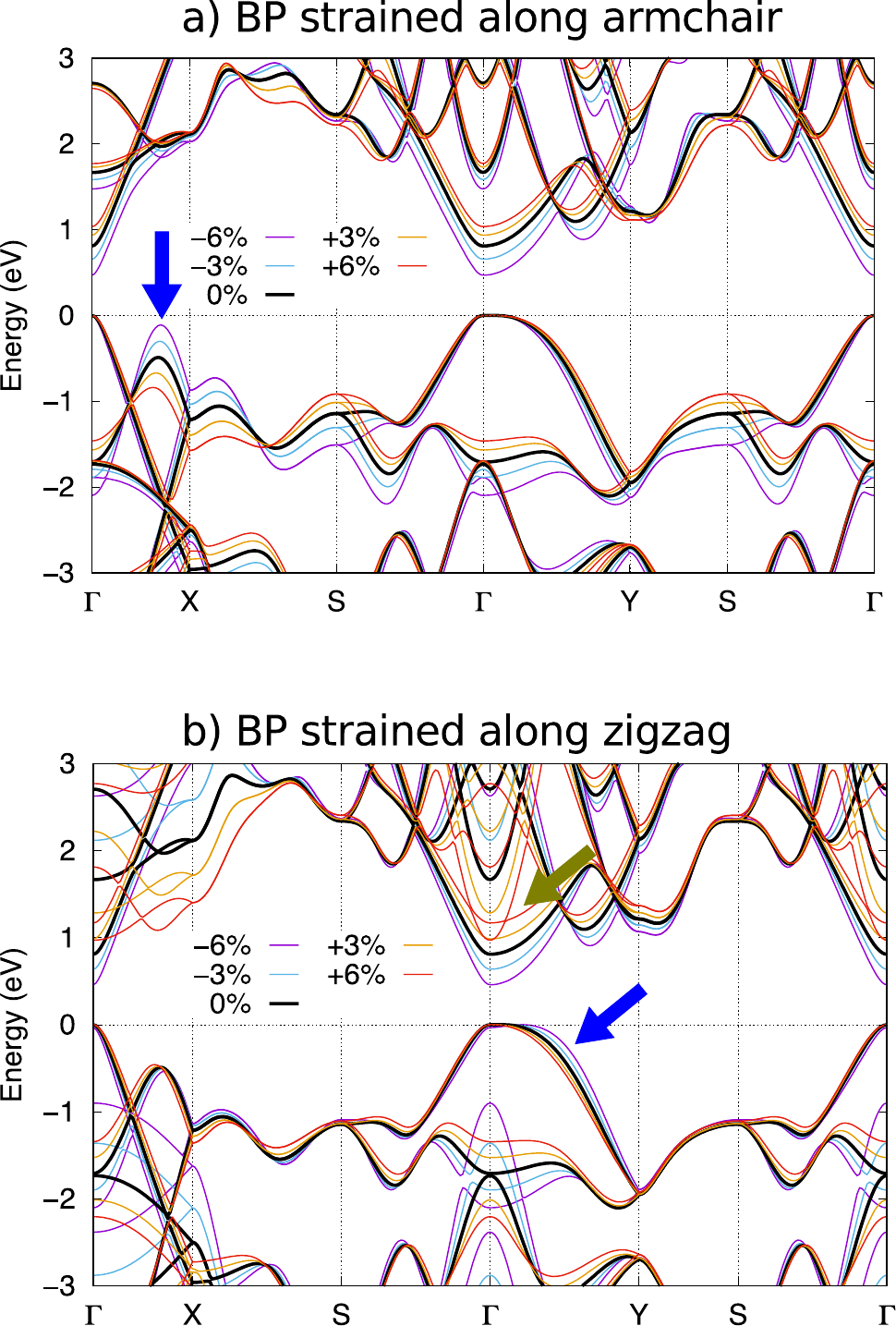}
    \caption{Band structure of the isolated BP monolayer. a) Strain ranging from -6\% to +6\% along the armchair axis; b) same strain applied on the zigzag axis. Blue and yellowish arrows highlight notable changes (see text).}
    \label{fig:bands_bp}
\end{figure}

\begin{figure}
    \centering
    \includegraphics[height=11 cm]{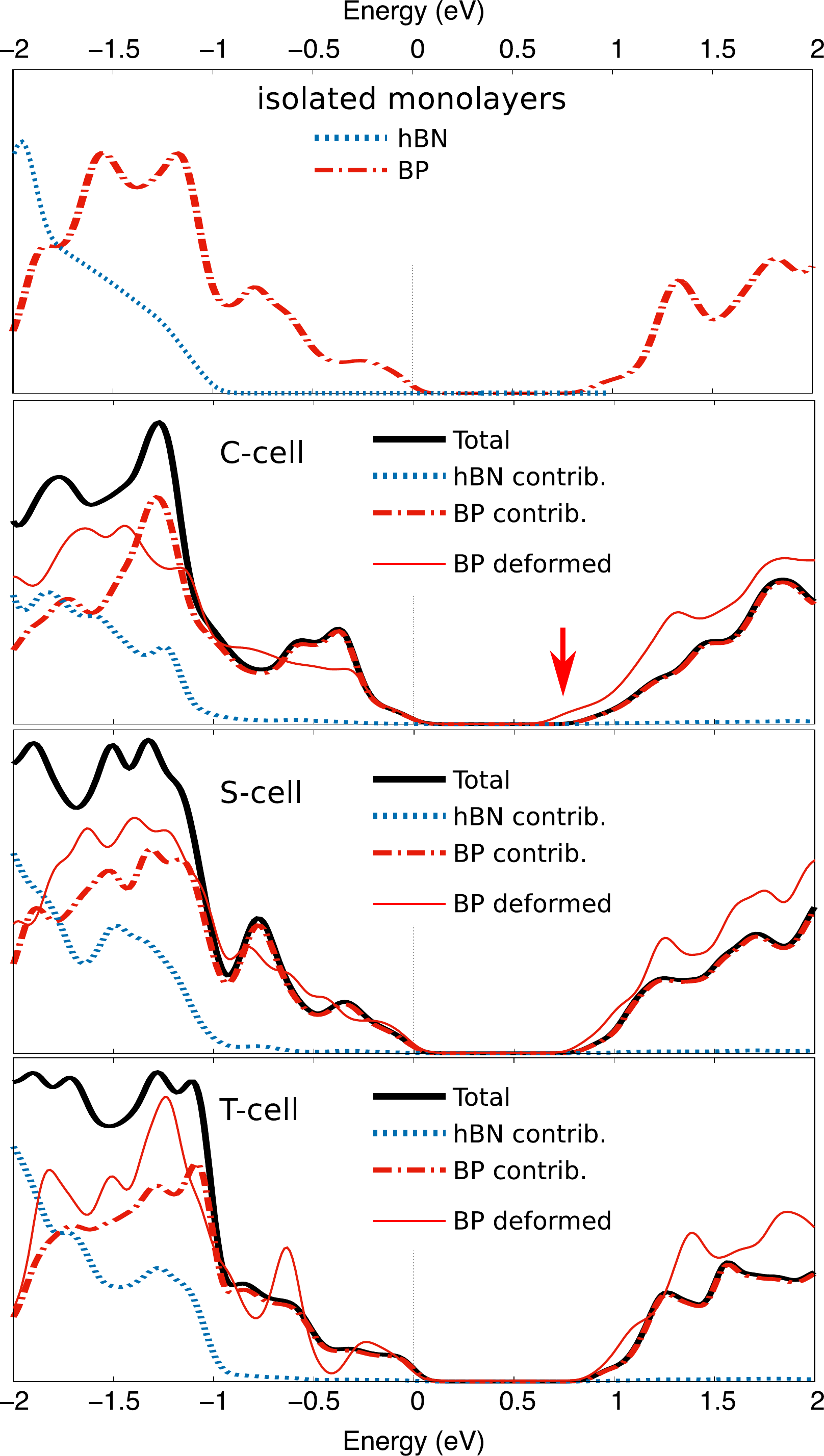}
    \caption{Density of states of the relaxed monolayers (top panel) and the three BP/hBN bilayers.
    In the bilayers, the dashed BP and the dotted hBN curves refer to the monolayer contributions to the total DOS, the thin red curve, is the deformed BP monolayer.
    The top valence of the bilayers and the BP monolayer are aligned at 0, while the top valence of the hBN monolayer is aligned approximately to the hBN contribution of the bilayers. All data have been broadened with a Gaussian of $\sigma=0.05$ eV.}
    \label{fig:dos_bphbn}
\end{figure}

Previous studies carried out on the C-cell~\cite{hu-hong_applmatinterf2015,steinkasserer-suhr-paulus_prb2016,zhang-wang-duan_chinphysb2016}, found that the top valence and the bottom conduction states were localised on the BP layer. In addition all these works report that the bilayer gap is between 0.2 and 0.07 eV larger than that of the isolated BP because of local electrostatic effects~\cite{steinkasserer-suhr-paulus_prb2016}.
We confirm that this is the case also in the other two cells.
As our PBE gap of the isolated BP monolayer is 0.81 eV (in quite good agreement with similar calculations~\cite{han_nanolett2014,steinkasserer-suhr-paulus_prb2016, hu-hong_applmatinterf2015,rasmussen_prb2016,cudazzo_prl2016}), we also confirm that there is a general trend of opening the gap in the bilayer configuration, even though not particularly in the C-cell.

We have shown that, in all BP/hBN bilayers, the BP sheet undergoes most of the deformations, preferentially along the armchair direction in the C-cell and the S-cell, and along the zigzag axis in the T-cell.
In Figure~\ref{fig:bands_bp} we report the band structure of the BP monolayer under uniaxial strain along the two crystallographic directions. 
Our results are consistent with the very complete study of Peng and coworkers~\cite{peng-wei-copple_prb2014}.
The band plot shows that contractions along the armchair axis may spoil the remarkable transport properties of BP because of the rise of an alternative top valence state in the $\Gamma-X$ (armchair) direction when strain is close to -6\% (blue arrow in Figure~\ref{fig:bands_bp}a).
This is the scenario we encountered in the C-cell and, in a lesser measure, in the S-cell.
Note that, despite the shearing angle, the features of the band structures that we discussed, look similar in the C-cell and the S-cell and evolve in a similar way. 
This consideration contradicts partially previous works on the subject, and suggests that BP/hBN bilayers where the zigzag axis of BP and hBN are aligned may lead to a degradation of the transport properties of BP because of the contact strain which leads to a substantial contraction of the armchair axis of BP.

Instead, in the T-cell most of the deformation is a dilatation along the zigzag axis of about 1.5\%.
Such a deformation has a minor impact on the armchair hole mobility since bands are quite unchanged along $\Gamma-X$. 
Instead the zigzag hole mobility may result slightly enhanced because the curvature of the top valence in $\Gamma-Y$ increases as the zigzag axis is expanded (blue arrow in Figure~\ref{fig:bands_bp}b).
On the other hand, at low degrees of tensile strain, the conduction band is a bit flattened, leading potentially to a lower electron mobility.
But, if the zigzag axis is highly extended (around 6\% of tensile strain), a band comes down from high-energy regions and becomes a new bottom conduction, as highlighted by the yellowish arrow.
Given the high dispersion of this new conduction state, electrons
should acquire a lighter effective mass and one should observe the electron mobility suddenly increase as the band crossing takes place.
To generalise this statement we claim that, if some twist angles allow one to force BP to stretch along the zigzag axis instead of contracting along the armchair axis, that would not only preserve the very good hole mobility along the armchair direction, but even lead to a substantial increase of the electron mobility in the zigzag direction on the proviso that the elongation is sufficiently important (around 6\%).

To characterize further the electronic properties of the three considered BP/hBN bilayers, we report the layer-projected density of states (LP-DOS) in Figure~\ref{fig:dos_bphbn}.
Thick black lines report the total DOS of the bilayers, while dotted blue and dashed-dotted red curves report the contribution of the hBN and the BP layer respectively.
In thin solid red, we also report the DOS of the isolated BP layer deformed as in the heterobilayer.

Comparing the gap of the structures, we identified a competition between the opening of the gap due to the interaction with the hBN and the closing because of the deformation.
This is true in all systems, but it is particularly evident in the C-cell where the two effects almost cancel (0.81 eV in the relaxed monolayer, 0.80 in the deformed one, 0.83 in the bilayer). 
This explains partially why we observed a much smaller opening of the gap with respect to the other works done on the C-cell configurations~\cite{han_nanolett2014,steinkasserer-suhr-paulus_prb2016, hu-hong_applmatinterf2015,rasmussen_prb2016,cudazzo_prl2016}.
A red arrow in Figure~\ref{fig:dos_bphbn} highlights this effect.

All total DOSs exhibit specific structures in the range between -1~eV and 0~eV.
Given the range of energy and the amplitude of these peaks, we expect similar features to be detectable by Scanning Tunneling Microscopy analysis, and possibly used as easy measures to characterise the stacking in future experiments.
However, these features are just vaguely related to the DOS of the deformed BP sheet, indicating that they cannot be put straightforwardly in relation to the deformation of the BP sheet.
Indeed, this indicates that the hBN layer does actually influence the electronic properties of the phosphorus layer beyond the simple application of some stress.

\subsection{The \Pblue/hBN bilayer}

\begin{figure}
    \centering
    \includegraphics[width=0.48\textwidth]{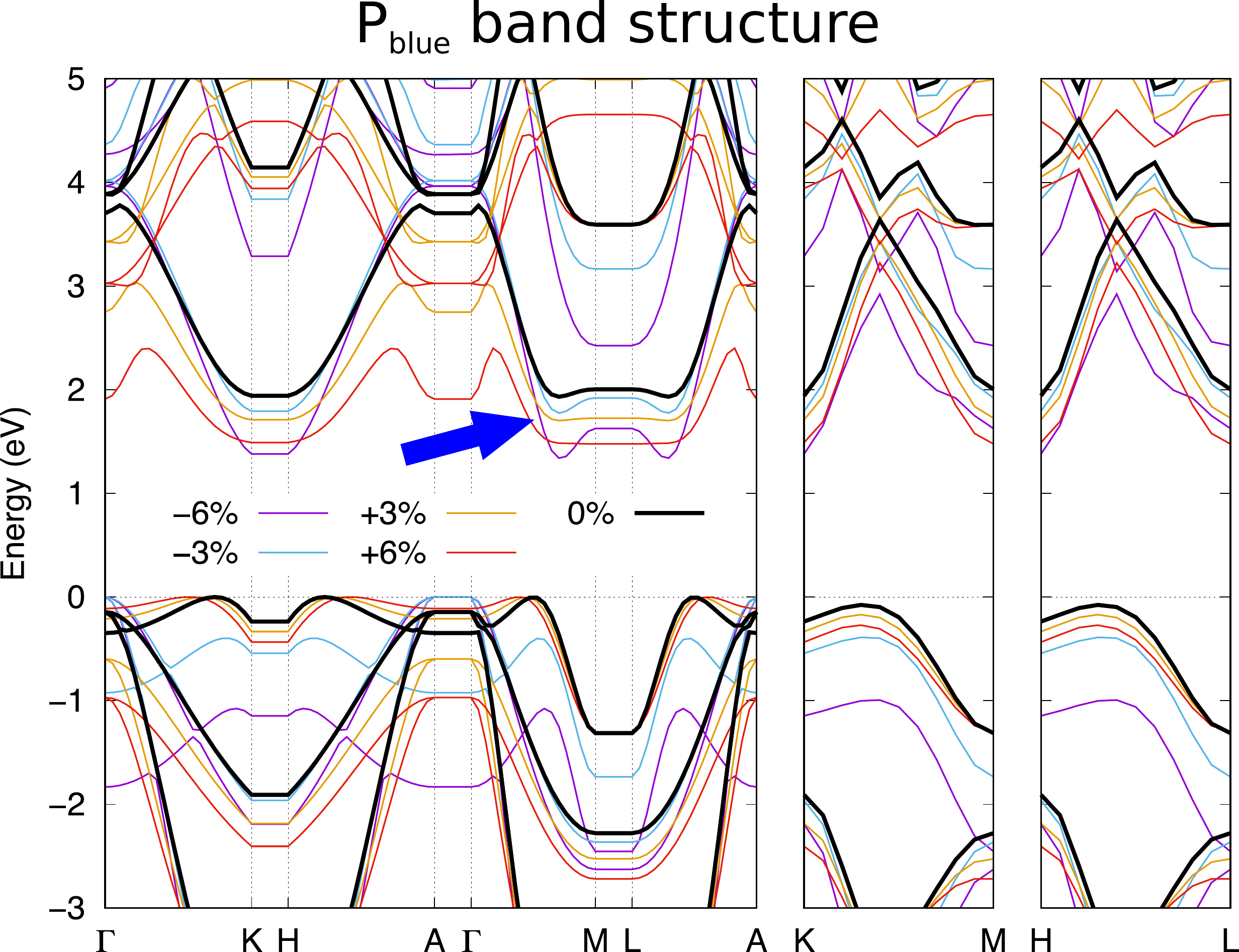}
    \caption{Band structure of the isolated \Pblue{ } monolayer biaxially strained.}
    \label{fig:bands_pblue}
\end{figure}

We now focus on the electronic properties and the LP-DOS of the \Pblue{}/hBN layer. As reported previously, the BN is slightly deformed (weak compression of 0.4\%) and does not lead to any particular electronic modifications.
Meanwhile, most of the deformation is sustained by the \Pblue{ }layer by a stretching of 1.25\% of the in-plane parameter and a reduction of the buckling height. If one refers to the band plot of the isolated \Pblue{ }monolayer under deformation (Figure~\ref{fig:bands_pblue}), one realises that stretching the \Pblue{ }layer of an amount of 1.25\% leads to a sizable flattening of the conduction band in the $\Gamma - M$ direction (blue arrow), and much weaker effects along $\Gamma - K$. This may have some impact in enhancing directional transport in this material because of the drastic increase of the electron effective mass along the $\Gamma - M$ axis.
This may also have some notable impact onto the optical properties of the layer. In particular, upon formation of electron-hole pairs, the flattening of the conduction band would result in a strong localization of the electronic part of the excitonic wavefunction. 

\begin{figure}
    \centering
    \includegraphics[height=7.3cm]{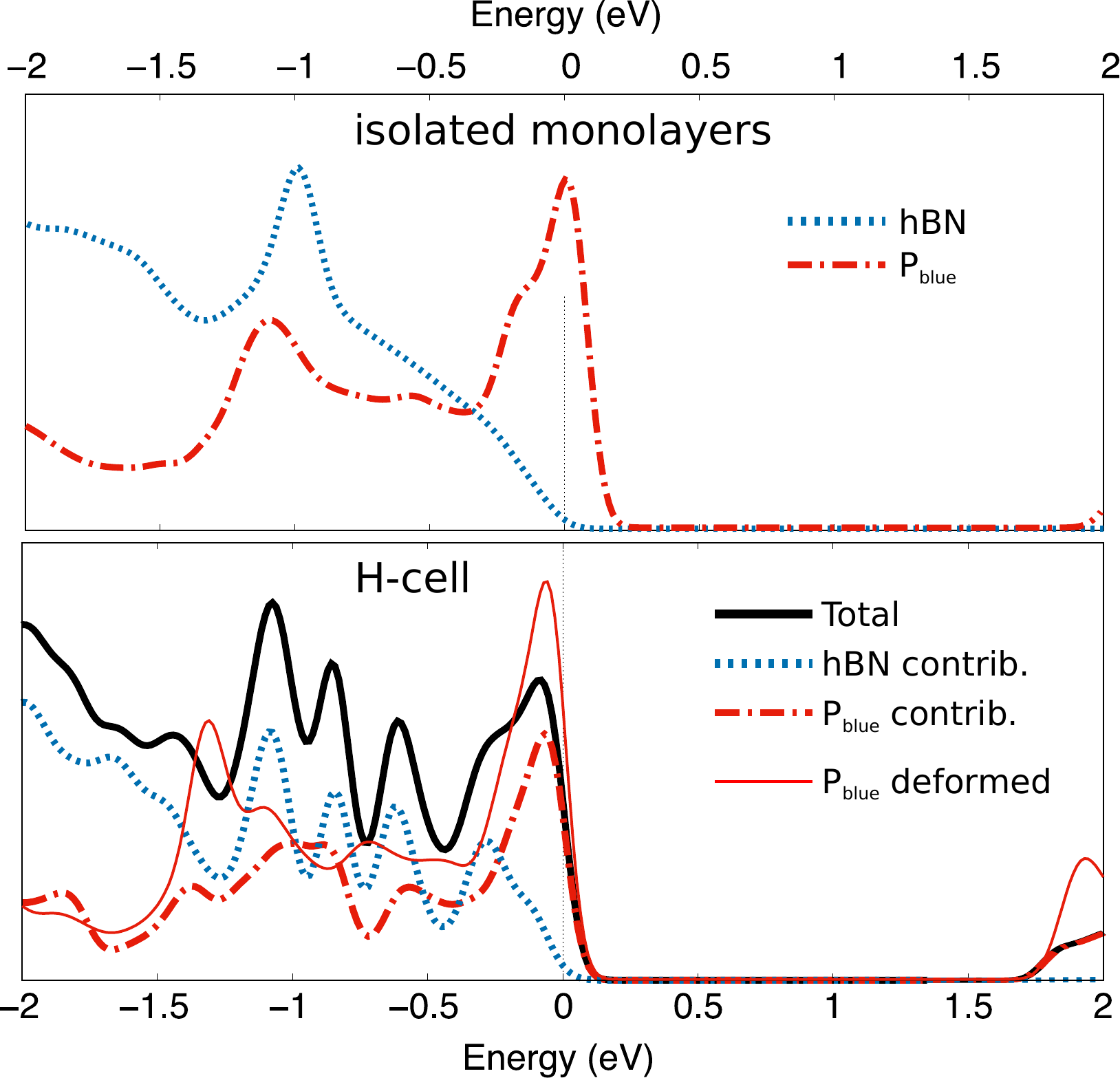}
    \caption{Density of states of the isolated hBN and \Pblue{ }monolayers (top panel) and the H-cell.
    In the bilayer (bottom panel), the \Pblue{ } and the hBN curves refer to the monolayer contributions to the total DOS.
    In red solid line, the deformed \Pblue{ } monolayer contribution.
    The top valence of all materials are aligned at 0.
    All data have been broadened with a Gaussian of $\sigma=0.05$ eV.}
    \label{fig:dos_pbluehbn}
\end{figure}

Considering the valence band, the LP-DOS reported in Figure~\ref{fig:dos_pbluehbn} shows an unexpected and striking feature. While the bottom conduction is completely localized on the \Pblue{ } layer, the top valence presents a non-negligible hybridisation between the two layers. 
From an excited-state perspective, this suggests that electron transport would take place only through the \Pblue{ } layer, while holes could hop from one layer to the other. Similarly, excitons would have a mixed character, being well localised in the \Pblue{ } layer for the electronic part of their wavefunction (especially if the \Pblue{ }is stretched, as discussed above), while presenting some charge-transfer characteristic in the hole part of it. This point is of particular interest because the localization of the electron and the hole into two different layers (interlayer exciton) suppresses most recombination channels, hence stabilizing the exciton even at room temperature leading to promising applications~\cite{withers_natmat2015,unucheck_nature2018}.

\section{Conclusions}

We have studied the structural and electronic properties of BP/hBN and \Pblue/hBN bilayers by means of DFT calculations with the intent of evaluating the impact of hBN on the phosphorus electronic properties in realistically relaxed structures.


In the BP/hBN systems, we find that there is a systematic repulsion between P and N atoms. Because of this, structures where the zigzag directions of BP and hBN are aligned (we called them C-cell structures) are not preferential configurations for the bilayer, at variance with those predicted in BP/graphene heterostructures~\cite{vantroeye_prm2018}.
Shearing the BP to reduce strain, has eventually a very high energetic cost and the resulting structure has even worse stability.
Instead, by misaligning the two layers (T-cell) through a twist angle it is possible to obtain configurations slightly more stable than the C-cell or, at worse, of comparable stability, with very low deformations of the BP layer.

The compressive strain along the armchair axis required to fit in the C-cell is expected to spoil the transport properties of BP because 
of a competition between the top valence in $\Gamma$ and a secondary valence band raising in between $\Gamma - X$. 
Instead,
upon a twist angle of 19$^\circ$, the BP layer mostly stretches along the zigzag axis leaving the band structure on the armchair axis almost unaltered.
As a result, the very promising charge mobility
of the BP layer is not spoiled, and may actually be enhanced upon stretches of about 6\%.

We also computed the stability, the strain distribution and the electronic structure of a \Pblue/hBN bilayer.
Also in this system, the phosphorus layer deforms more than the hBN (biaxial dilatation of 1.25\%), even though the deformation energy is shared more equally between the two layers as a consequence of the higher rigidity of the \Pblue{ }with respect of that of the BP.
We observed that, at low compression, a flat band develops in the bottom conduction and that there is a non-negligible hybridization between the valence states of the two layers.
These observations imply that (i) the electron mobility of \Pblue{ } can be suppressed and the electronic part of excitonic wavefunctions can be localised on the \Pblue{ }layer by relatively low strain; and (ii) that, upon creation of electron-hole pairs, holes can hop from one layer to the other.

\medskip

This work received funding from the European Union’s Horizon 2020 research and innovation programme under grant agreements No. 785219 (Graphene Core 2) and from the French national research agency (ANR) under the grant agreement No. ANR-17-CE24-0023-01 (EPOS-BP).
%

%
%
%

\end{document}